\documentclass[epj
]{svjour}
\usepackage{graphics}
\begin{document}
\titlerunning{Derivation of Non-Local Macroscopic Traffic Equations  
from Microscopic Car-Following Models}
\title{Derivation of Non-Local Macroscopic Traffic Equations and Consistent Traffic Pressures 
from Microscopic Car-Following Models}
\author{Dirk Helbing
}                     
%
%
\institute{ETH Zurich, UNO D11, Universit\"atstr. 41, 8092 Zurich, Switzerland}
\date{Received: date / Revised version: date}
%
\abstract{
This contribution compares several different approaches allowing one to derive macroscopic traffic equation directly from microscopic car-following models. While it is shown that some conventional approaches lead to theoretical problems, it is proposed to use a smooth particle hydrodynamic approach and to avoid gradient expansions. The derivation circumvents approximations and, therefore, demonstrates the large range of validity of macroscopic traffic equations, without the need of averaging over many vehicles. It also gives an expression for the ``traffic pressure'', which generalizes previously used formulas. Furthermore, the method avoids theoretical inconsistencies of macroscopic traffic models, which have been criticized in the past by Daganzo and others.
} 
\maketitle
\section{Introduction}

In order to describe the dynamics of traffic flows, a large number of mathematical models has been developed. The analysis of the spatio-temporal features and statistics of traffic patterns has often been done with methods from statistical physics and non-linear dynamics. An overview of modeling approaches and methods is, for example, given in Refs. \cite{Schadschneider,Review,Nagatani,Nagel}, among them cellular automata, ``microscopic'' car-following models, ``mesoscopic'' gas-kinetic, and macroscopic traffic models. 
\par
Cellular automata can often be interpreted as discretized versions of car-following models, while gas-kinetic models have frequently been used to derive macroscopic from microscopic models. Such derivations were driven by the desire to improve phenomenological specifications of macroscopic traffic models \cite{Kuehne,Papageorgiou,Kerner93}, which were criticized to have unrealistic properties \cite{Daganzo}. However, the derivation of gas-kinetic models from car-following models usually simplifies the interactions among vehicles by a collisional approach assuming immediate braking maneuvers. Moreover, the derivation of macroscopic traffic models from gas-kinetic ones terminates an infinite and poorly converging
series expansion, which replaces dynamical equations for higher moments of the velocity distribution by simplified equilibrium relationships \cite{granular}. 
\par
Although this leads to macroscopic equations which work well in most theoretical and practical aspects \cite{GKT}, the implications of the approximations are hardly known. Moreover, the approach seems to require an averaging over at least 100 vehicles for each speed class and spatial location. While this constitutes no problem for gases with $10^{23}$ particles within a small volume, for traffic flows this would require an averaging over spatial intervals much greater than the scale on which traffic flow changes. Hence, it is not well understood, whether or why macroscopic traffic equations can be used at all.
\par
In this paper, we will therefore focus on attempts to derive macroscopic traffic equations directly from microscopic ones. Doing so, we will compare three different approaches: First, we study the gradient expansion approach in Sec. \ref{Gradient}. Second, we turn to the linear interpolation approach in Sec. \ref{Interpolation}. Third, we discuss the smooth particle hydrodynamics approach in Sec. \ref{Smooth} and compare the results with macroscopic traffic models such as the Payne model, the Aw-Rascle model, and a non-local traffic model. In the Conclusions, we summarize and discuss our results, in particular with regard to the mathematical form of the traffic pressure and the theoretical consistency of macroscopic traffic models.

\section{The Gradient Expansion Approach} \label{Gradient}

Already in the 1970's, Payne \cite{Payne1,Payne2} used a gradient expansion approach to derive a macroscopic velocity equation complementing the continuity equation 
\begin{equation}
\frac{\partial \rho}{\partial t} + \frac{\partial }{\partial x} \Big[ \rho(x,t)V(x,t) \Big]  = 0 \, .
\label{CONT}
\end{equation}
It relates the vehicle density $\rho(x,t)$ at location $x$ and time $t$ with the average velocity $V(x,t)$ or the vehicle flow 
\begin{equation}
Q(x,t) = \rho(x,t)V(x,t) \, ,
\end{equation}
respectively, and describes the conservation of the number of vehicles \cite{Whitham}. 
\par
Payne derived his model from Newell's car-following model \cite{Newell}
\begin{equation}
v_i(t+\tau) = v_{\rm o}\big(d_i(t)\big) \, ,
\label{Newell}
\end{equation}
which assumes that the speed $v_i(t)$ of vehicle $i$ at time $t$ will be adjusted with a delay of $\tau$ to some optimal speed $v_{\rm o}$, which depends on the distance $d_i(t) = x_{i-1}(t)-x_i(t)$ between the location of the leading vehicle $x_{i-1}(t)$ and the location $x_i(t)$ of the following car.
\par
Payne identified microscopic and macroscopic velocities as follows:
\begin{eqnarray}
 v_i(t+\tau) &=& V(x+V\,\tau, t+\tau) \nonumber \\
 &\approx&   V(x,t) + V\,\tau\, \frac{\partial V(x,t)}{\partial x} +
 \tau \, \frac{\partial V(x,t)}{\partial t} \, . \qquad 
\end{eqnarray}
Then, Taylor approximations (gradient expansions) were used in several places.
For example, Payne substituted the inverse of the distance $d_i$
to the leading vehicle by the density $\rho$ at the place
$x + d_i(t)/2$ in the middle between the leading and the
following vehicle. In this way, he obtained
\begin{eqnarray}
 \frac{1}{d_i(t)} &=& \rho\left(x+ \frac{d_i(t)}{2},t\right) 
 = \rho\left( x + \frac{1}{2\rho},t \right) \nonumber \\
 &\approx & \rho(x,t) + \frac{1}{2\rho} \frac{\partial
 \rho(x,t)}{\partial x}  \, .
 \label{obta}
\end{eqnarray}
When defining the so-called equilibrium velocity $V_{\rm e}(\rho)$ through
\begin{equation}
 V_{\rm e}(\rho) = v_{\rm o}\left(\frac{1}{\rho}\right) \qquad \mbox{or} \qquad V_{\rm e}\left(\frac{1}{d_i}\right) = v_{\rm o}(d_i) \, ,
\end{equation}
a first order Taylor approximation and Eq. (\ref{obta}) imply
\begin{eqnarray}
 v_{\rm o}\big(d_i(t)\big) &=& V_{\rm e}\left(\frac{1}{d_i(t)}\right) \nonumber \\ 
 &\approx & V_{\rm e}\big(\rho(x,t)\big) + \frac{1}{2\rho(x,t)} \, \frac{dV_{\rm e}(\rho)}{d\rho} \,
 \frac{\partial \rho(x,t)}{\partial x} \, . \qquad
\end{eqnarray}
Starting from the previous equations,
one finally arrives at Payne's macroscopic velocity equation
\begin{equation}
 \frac{\partial V}{\partial t} + V \frac{\partial V}{\partial x}
 =  \frac{1}{\tau} \left[ V_{\rm e}(\rho) - \frac{D(\rho)}{\rho}
 \frac{\partial \rho}{\partial x}- V(x,t) \right] \, ,
\label{payne}
\end{equation}
where we have introduced the density-dependent diffusion
\begin{equation}
 D(\rho) = - \frac{1}{2} \frac{dV_{\rm e}(\rho)}{\partial \rho}
 = \frac{1}{2} \left| \frac{d V_{\rm e}(\rho)}{d \rho} \right| \ge  0 \, .
\end{equation}
The single terms of Eq.~(\ref{payne}) have the following
interpretation:
The term $V\partial V/\partial x$ is called the {\em transport term} and describes a motion of the velocity profile with the vehicles. 
The term $-[D(\rho)/(\rho \, \Delta t)]\partial \rho /\partial x$
is called {\em anticipation term}, as it reflects the reaction of drivers to the traffic situation in front of them.
The {\em relaxation term} $[V_{\rm e}(\rho)-V]/\Delta t$ delineates the
adaptation of the average velocity $V(x,t)$ to the density-dependent
{\em equilibrium velocity} $V_{\rm e}(\rho)$ with a delay $\tau$. 
\par
Other authors have applied similar gradient expansions to the optimal velocity model
defined by
\begin{equation}
\frac{dv_i(t)}{dt} = \frac{1}{\tau} \Big[ v_{\rm o}\big(d_i(t)\big) - v_i(t) \Big] 
\label{Bando}
\end{equation}
with $dd_i/dt = v_{i-1}(t) - v_i(t)$, see e.g. Refs. \cite{Berg,Kims}.
Equation (\ref{Bando}) results from the Newell model (\ref{Newell}) by a first-order Taylor approximation
$v_i(t+\tau) \approx v_i(t) + \tau \, dv_i/dt$. Regarding the derivation of macroscopic traffic equations from the optimal velocity model, it is also worth reading Refs. \cite{Berg,Kims}.
\par
One weakness of the gradient expansion approach is that it implicitly assumes small gradients in order to be mathematically valid. It is well-known, however, that many microscopic and macroscopic traffic equations give rise to emergent traffic jams, which are related with steep gradients. That would require the consideration of higher-order terms and lead to macroscopic traffic equations that are not anymore simple and well tractable (even numerically). Let us, therefore, study other approaches to determine macroscopic from microscopic equations.

\section{The Linear Interpolation Approach} \label{Interpolation}

The optimal velocity model may be also written in the form
\begin{equation}
 \frac{dv_i}{dt} = a_i(t) = \frac{v^0 - v_i(t)}{\tau} + f \big(d_i(t)\big) \, ,
 \label{acceq}
\end{equation}
where $a_i(t)$ denotes the acceleration, $v^0$ the ``desired velocity'' or ``free speed'', and
\begin{equation}
f(d_i) = \frac{v_{\rm o}(d_i) -v^0}{\tau} \le 0
\end{equation}
the repulsive interaction among the leading vehicle $i-1$ and its follower $i$.
\par
In Ref. \cite{MCM}, it has been suggested to establish a micro-macro link between microscopic and macroscopic traffic variables by the definitions
\begin{eqnarray}
 \rho(x,t) &=& \frac{\displaystyle\frac{1}{x_{i}(t)-x_{i+1}(t)} \big[ x_{i-1}(t) - x \big]}{x_{i-1}(t) - x_i(t)}  \nonumber \\
 &+& \frac{\displaystyle\frac{1}{x_{i-1}(t)-x_i(t)}\big[x - x_i(t)\big]} {x_{i-1}(t) - x_i(t)} \, ,  \qquad \\[2mm]
 V(x,t) &=& \frac{v_i(t) \big[ x_{i-1}(t) - x \big] + v_{i-1}(t) \big[x - x_i(t)\big]}
 {x_{i-1}(t) - x_i(t)} \, , \label{VD} \\[2mm]
 A(x,t) &=& \frac{a_i(t) \big[ x_{i-1}(t) - x \big] + a_{i-1}(t) \big[x - x_i(t)\big]} 
 {x_{i-1}(t) - x_i(t)} \, . \qquad 
\end{eqnarray}
These definitions assume that the macroscopic variables in the vehicle locations $x=x_i(t)$ would be given by the microscopic ones, while in locations $x$ {\it between} two vehicles, they would be defined by linear interpolation. 
\par
Let us consider the consequences of such an approach.
For this, we determine the partial derivative of 
\begin{equation}
G(x,t) = \frac{g_i(t) \big[ x_{i-1}(t) - x \big] + g_{i-1}(t) \big[x - x_i(t)\big]}
 {x_{i-1}(t) - x_i(t)} 
 \label{linint}
\end{equation}
with respect to $x$, which gives
\begin{equation}
\frac{\partial G(x,t)}{\partial x} = \frac{-g_i(t)  + g_{i-1}(t)}
 {x_{i-1}(t) - x_i(t)} 
\end{equation}
for any specification of $g_i(t)$, for example, $g_i(t) =v_i(t)$. The partial derivative with respect to time is
\begin{eqnarray}
\frac{\partial G(x,t)}{\partial t} &=& \frac{\frac{dg_i(t)}{dt} \big[ x_{i-1}(t) - x \big] 
+ g_i(t) \frac{dx_{i-1}(t)}{dt} } {x_{i-1}(t) - x_i(t)} \nonumber \\
&+& \frac{\frac{dg_{i-1}(t)}{dt} \big[x - x_i(t)\big] - g_{i-1}(t) \frac{dx_i(t)}{dt} } 
 {x_{i-1}(t) - x_i(t)} \nonumber \\[2mm]
 &-&  \frac{\left( \frac{dx_{i-1}(t)}{dt} - \frac{dx_i(t)}{dt} \right) 
  g_i(t) \big[ x_{i-1}(t) - x \big]  }
 {\big[x_{i-1}(t) - x_i(t)\big]^2} \nonumber \\
&-&  \frac{\left( \frac{dx_{i-1}(t)}{dt} - \frac{dx_i(t)}{dt} \right) 
  g_{i-1}(t) \big[x - x_i(t)\big] }
 {\big[x_{i-1}(t) - x_i(t)\big]^2} \, .\qquad 
\end{eqnarray}
For $g_i(t) = v_i(t) = dx_i/dt$ and with $dv_i/dt = a_i(t)$, this formula simplifies to the following expression:
\begin{eqnarray}
\frac{\partial V(x,t)}{\partial t} &=& \frac{a_i(t) \big[ x_{i-1}(t) - x \big] 
+ v_i(t) v_{i-1}(t)}{x_{i-1}(t) - x_i(t)}  \nonumber \\
&+& \frac{a_{i-1}(t) \big[x - x_i(t)\big] - v_{i-1}(t) v_i(t) } 
 {x_{i-1}(t) - x_i(t)} \nonumber \\
 &-&  \frac{v_{i-1}(t) - v_i(t)}{ x_{i-1}(t) - x_i(t) }\nonumber \\
 & & \times \frac{v_i(t) \big[ x_{i-1}(t) - x \big] + v_{i-1}(t) \big[x - x_i(t)\big] }
 {x_{i-1}(t) - x_i(t)} \nonumber \\[2mm]
  &=& A(x,t) - \frac{\partial V(x,t)}{\partial x} V(x,t) \, .
\end{eqnarray}
As a consequence, we find the exact relationship
\begin{equation}
 \frac{\partial V(x,t)}{\partial t} + V(x,t) \frac{\partial V(x,t)}{\partial x} = A(x,t) \, .
\end{equation}
This would be fully compatible with Payne's macroscopic traffic equation (\ref{payne}), if
\begin{equation}
 A(x,t) = \frac{1}{\tau} \Big[ V_{\rm e}(\rho) - V(x,t) \Big] - \frac{D(\rho)}{\tau \rho(x,t)}
 \frac{\partial \rho}{\partial x} \, .
\end{equation}
However, the expression for $g_i(t) = 1/[x_{i-1}(t)-x_i(t)]$ does not simplify in a way 
that would finally lead to the continuity equation (\ref{CONT}). Therefore, a micro-macro link based on the linear interpolation (\ref{linint}) of the microscopic variables $g_i(t)$ does not exactly imply the conservation of the number of vehicles, i.e. it is theoretically not consistent. Nevertheless, it works surprisingly well in practise \cite{MCM}.

\section{The Smooth Particle Hydrodynamics Approach} \label{Smooth}

\subsection{Derivation of the Continuity Equation}

In this section, we will derive macroscopic traffic equations directly from microscopic ones. 
We will start with the derivation of the continuity equation from the equation of motion $dx_i/dt = v_i$,
using a ``trick'' that I learned from Isaac Goldhirsch. For this,
we represent the location $x_i(t)$ of an element $i$ in space by a
delta function $\delta(x - x_i(t))$, which may be treated here like a very narrow Gaussian distribution.
Moreover, we introduce a {\it symmetrical} smoothing function
\begin{equation}
s(x'-x) = s(|x'-x|) = s(x-x') \, , 
\end{equation}
for example, a Gaussian distribution with a finite variance or a differentiable approximation of a triangular function or a rectangular one. The smoothing function shall be normalized by demanding
\begin{equation}
 \int\limits_{-\infty}^\infty d x' \; s(x'-x) = 1
\end{equation}
for any value of $x$. With this, we define the local density
\begin{eqnarray}
 \rho(x,t) &=& \int\limits_{-\infty}^\infty dx' \; s(x'-x) 
 \sum_i \delta \big(x' - x_i(t)\big) \label{densdef} \\
&=& \sum_i s(x_i(t) - x) \, . \label{densdef2}
\end{eqnarray}
Herein, we sum up over all particles $i$. Note that the replacement of the conventional formula $\sum_i \delta(x_i(t) - x)$ for the vehicle density by the formula $\sum_i s(x_i(t) - x)$ corresponds to a substitution of point-like particles by ``fuzzy'' particles, which is the idea behind smooth particle hydrodynamics.
\par
Now, we define the average velocity $V(x,t)$ as usual via a weighted average
with the weight function $\delta (x' - x_i(t)) s(x'-x)$:
\begin{eqnarray}
 V(x,t) &=& \frac{\int\limits_{-\infty}^\infty d x' \; {\displaystyle \sum_i} v_i(t) \delta \big(x' - x_i(t)\big) s(x'-x)}
{\int\limits_{-\infty}^\infty d x' \; {\displaystyle \sum_i} \delta \big(x - x_i(t)\big) s(x'-x)} \nonumber \\
&=& \frac{\int\limits_{-\infty}^\infty d x' \; {\displaystyle \sum_i} v_i(t) \delta \big(x' - x_i(t)\big) s(x'-x)}
{\rho(x,t)} \label{veldef} \nonumber \\[3mm]
&=&\frac{\displaystyle\sum_i v_i(t) s(x_i(t) - x)}{\displaystyle\sum_i s(x_i(t) - x)} \nonumber \\
&=&  \frac{\displaystyle\sum_i v_i(t) s(x_i(t) - x)}{\rho(x,t)} \, .
\label{veldef2}
\end{eqnarray}
This implies the well-known fluid-dynamic flow relationship
\begin{equation}
 Q(x,t) = \rho(x,t)V(x,t)  \, . 
\end{equation}
Differentiation of Eq. (\ref{densdef}) with respect to time and application of the chain rule gives
\begin{eqnarray}
& &  \frac{\partial \rho(x,t)}{\partial t} \nonumber \\
&=& \int\limits_{-\infty}^\infty d x' \; \sum_i \left( -\frac{dx_i}{dt}\right) \cdot
\left[ \frac{\partial}{\partial x'} \delta \Big(x' - x_i(t)\Big)\right] s(x'-x) \nonumber \\
&=& \int\limits_{-\infty}^\infty d x' \; \sum_i v_i(t) 
\delta \Big(x' - x_i(t)\Big)\left[ \frac{\partial}{\partial x'} s(x'-x) \right] \, , \label{contder}
\end{eqnarray}
where we have applied partial integration to obtain the last results. That is, we have used the theorem
\begin{eqnarray}
& &  \int\limits_{-\infty}^\infty d x' \; \left[ \frac{\partial}{\partial x'} u(x')\right] v(x') \nonumber \\
&=&  \Big[ u(x)v(x)\Big]_{-\infty}^\infty
 - \int\limits_{-\infty}^\infty u(x')\left[\frac{\partial}{\partial x'}  v(x')\right] \, ,
\end{eqnarray}
considering the vanishing of the first term after the equality sign due to the vanishing of $u(x)v(x)$ at the boundaries. Taking into account the symmetry of
the smoothing function $s(x'-x)$, we may replace $\partial s(x'-x)/\partial x'$
by $-\partial s(x'-x)/\partial x$, which finally yields Eq. (\ref{CONT}) as follows:
\begin{eqnarray}
 \frac{\partial \rho(x,t)}{\partial t} &=& - \frac{\partial}{\partial x} \int\limits_{-\infty}^\infty dx' \; 
 \sum_i v_i(t) \delta \Big(x' - x_i(t)\Big) s(x'-x)  \nonumber \\
&=& - \frac{\partial}{\partial x} \Big[ \rho(x,t)V(x,t) \Big] \, .
\label{bis}
\end{eqnarray}
To obtain this desired result, we have finally
applied the definition (\ref{veldef}) of the average velocity $V(x,t)$. As a consequence of this, the validity of the continuity equation does not require an averaging over large numbers of entities, i.e.
macroscopic volumes to average over. This makes the equation absolutely fundamental and explains its large range of validity.

\subsection{Derivation of the Macroscopic Velocity Equation}

In order to derive the equation for the average velocity, we start by deriving the formula
\begin{equation}
 \rho(x,t)V(x,t)  = \sum_i v_i(t) s\big(x_i(t) - x\big)
\end{equation}
for the vehicle flow with respect to time. This gives
\begin{eqnarray}
\frac{\partial}{\partial t} \big[ \rho(x,t)V(x,t) \big] 
&=&   \sum_i \frac{dv_i(t)}{dt} s\big(x_i(t)- x\big) \nonumber \\
&+&  \sum_i v_i(t) \frac{\partial}{\partial x_i} \Big[ s\big(x_i(t)-x\big) \Big] \frac{dx_i(t)}{dt} \nonumber \\
&=&   \sum_i a_i(t) s\big(x_i(t)- x\big) \nonumber \\
&-&  \frac{\partial}{\partial x} \sum_i [v_i(t)]^2 \Big[ s\big(x_i(t)-x\big) \Big] .\qquad 
\end{eqnarray}
Introducing $\delta v_i(x,t) = v_i(t) - V(x,t)$ and defining the velocity variance
\begin{eqnarray}
 \theta(x,t) &=& \frac{\int\limits_{-\infty}^\infty  d x' \; \sum_i [v_i(t) -V(x,t)]^2 
 \delta \big(x' - x_i(t)\big) s(x'-x)}{\int\limits_{-\infty}^\infty  d x' \; \sum_i 
 \delta \big(x' - x_i(t)\big) s(x'-x)} \label{vardef} \nonumber \\
&=& \frac{ \sum_i [v_i(t) -V(x,t)]^2 s(x_i(t)-x)}{\sum_i s(x_i(t) - x)} \nonumber \\[2mm]
 &=& \frac{\sum_i [\delta v_i(x,t)]^2 s(x_i(t) - x)}{\rho(x,t)} \qquad 
\label{vardef2}
\end{eqnarray}
similarly to the average velocity (\ref{veldef2}), we can make the decomposition
\begin{eqnarray}
& &\sum_i [v_i(t)]^2  s\big(x_i(t)-x\big) \nonumber \\
&=& \sum_i [V(x,t) + \delta v_i(x,t)]^2  s\big(x_i(t)-x\big) \nonumber \\
&=& \sum_i \Big\{ [V(x,t)]^2 + 2 V(x,t)\delta v_i(x,t) \nonumber \\
& & + [\delta v_i(x,t)]^2 \Big\}  s\big(x_i(t)-x\big) \nonumber \\
&=& \rho(x,t) [V(x,t)]^2 + 2\rho(x,t)V(x,t)\big[V(x,t) - V(x,t)\big] \nonumber \\[2mm]
& & + \rho(x,t) \theta(x,t) \, , \qquad 
\end{eqnarray}
where we have considered
\begin{eqnarray}
& &  \sum_i  \delta v_i(x,t)  s\big(x_i(t)-x\big) \nonumber \\
&=& \sum_i  \Big[ v_i(t) - V(x,t) \Big]  s\big(x_i(t)-x\big) \nonumber \\
&=& Q(x,t) - \rho(x,t) V(x,t) = 0 \, ,
\end{eqnarray} 
see Eqs. (\ref{veldef2}) and (\ref{densdef2}).
Altogether, we get
\begin{eqnarray}
\frac{\partial}{\partial t} \big[ \rho(x,t)V(x,t) \big] 
&=& - \frac{\partial}{\partial x} \Big\{ \rho(x,t)\big[V(x,t)^2 + \theta(x,t)\big]\Big\} \nonumber \\
&+&  \sum_i a_i(t) s\big(x_i(t)- x\big) \, .  \label{whatever}
\end{eqnarray}
Now, we carry out the partial differentiation applying the product rule of Calculus.
Taking into account
\begin{equation}
\rho(x,t) \frac{\partial V(x,t)}{\partial t} = 
- V(x,t) \frac{\partial \rho(x,t)}{\partial t}
+ \frac{\partial}{\partial t} \big[ \rho(x,t)V(x,t) \big]
\end{equation}
and
\begin{eqnarray}
& & \frac{\partial}{\partial x} \Big\{ \big[ \rho(x,t)V(x,t)] V(x,t) \Big\} \nonumber \\
&=& \rho(x,t) V(x,t) \frac{\partial V}{\partial x} \nonumber \\
&+& V(x,t) \frac{\partial}{\partial x} \Big[ \rho(x,t)V(x,t) \Big]
\, , 
\end{eqnarray}
we obtain with Eq. (\ref{whatever})
\begin{eqnarray}
& & \rho(x,t) \frac{\partial V(x,t)}{\partial t} \nonumber \\
&=& - V(x,t) \frac{\partial \rho(x,t)}{\partial t} 
- V(x,t) \frac{\partial}{\partial x} \big[ \rho(x,t)V(x,t)\big]  \nonumber \\ 
&-& \rho(x,t)V(x,t) \frac{\partial V(x,t)}{\partial x} - \frac{\partial }{\partial x} \big[ \rho(x,t)\theta(x,t)\big] \nonumber \\
&+&  \sum_i a_i(t) s\big(x_i(t)- x\big) \, . 
\end{eqnarray}
Inserting the continuity equation (\ref{bis}) for $\partial \rho/\partial t$ and dividing the above
equation by $\rho(x,t)$ finally gives the velocity equation
\begin{eqnarray}
& & \frac{\partial V(x,t)}{\partial t} + V(x,t) \frac{\partial V(x,t)}{\partial x}  \nonumber \\
&=& - \frac{1}{\rho(x,t)}\frac{\partial }{\partial x} \big[ \rho(x,t)\theta(x,t)\big] \nonumber \\
&+&  \frac{1}{\rho(x,t)} \sum_i a_i(t) s\big(x_i(t)- x\big) \, .
\end{eqnarray}
Inserting Eq. (\ref{acceq}) for $a_i(t)$, we find
\begin{eqnarray}
& &  \sum_i a_i(t) s\big(x_i(t)- x\big) \nonumber \\
&=& \sum_i \left[ \frac{v^0 - v_i}{\tau} + \sum_i f\big(d_i(t)\big)
\right] s\big(x_i(t)- x\big) \nonumber \\
&=& \frac{v^0 - V(x,t)}{\tau} + \sum_i f\big(d_i(t)\big) s\big(x_i(t)- x\big) \, .
\label{reve}
\end{eqnarray}
For further simplification, let us now specify the smoothing function by the rectangular
function
\begin{equation}
s(x_i - x) = \frac{\varrho}{2} \cdot \left\{
\begin{array}{ll}
1 & \mbox{if } |x_i - x| \le 1/\varrho \\
0 & \mbox{otherwise,}
\end{array}
\right. 
\end{equation}
with a large enough smoothing window of length $\Delta x = 2/\varrho$. Then, the number of vehicles $i$ within the
smoothing interval $[x-1/\varrho,x+1/\varrho]$ is expected to be $\rho \, \Delta x = 2 \rho/\varrho$, where
$\rho$ represents the average vehicle density in this interval. Therefore,
\begin{equation}
 \rho(x,t) = \sum_i s\big(x_i(t)-x\big) = \frac{2\rho}{\varrho} \frac{\varrho}{2} = \rho \, ,
\end{equation}
which shows the consistency of this approach.
\par
If the smoothing parameter
$\varrho$ is specified via the inverse vehicle distance
\begin{equation}
 \varrho = \varrho_k = \frac{1}{d_k} = \frac{1}{x_{k-1} - x_{k}} = \rho(x,t) 
 \quad \mbox{for} \quad x_{k} < x \le x_{k-1},
 \label{Using}
\end{equation} 
the smoothing window of length $\Delta x = 2/\varrho$ will usually contain only
two vehicles $k-1$ and $k$ with $x_{k} \le x \le x_{k-1}$. With this, the sum over $i$ 
reduces to two terms with $i=k$ and $i=k-1$ only. This finally yields 
\begin{eqnarray}
V(x,t) &=& \sum_i v_i(t) s(x_i(t) - x) \nonumber \\
&=& v_k(t) s(x_{k}(t)-x)   + v_{k-1}(t) s(x_{k-1}(t)-x) \nonumber \\
&=& \frac{\varrho}{2} \big[ v_{k-1}(t) + v_{k}(t) \big] \nonumber \\
&=& \rho(x,t) \frac{v_{k-1}(t) + v_{k}(t)}{2} 
\end{eqnarray}
and, considering Eq. (\ref{Using}),  
\begin{eqnarray}
\sum_i s(x_i(t)-x) f\big(d_i(t)\big) &=& \frac{\varrho}{2} f(d_k) + \frac{\varrho}{2}f(d_{k-1}) \nonumber \\
&=& \frac{\varrho}{2} f\left(\frac{1}{\varrho_k}\right) + \frac{\varrho}{2}f\left(\frac{1}{\varrho_{k-1}}\right) \nonumber \\ 
&=& \frac{\rho(x,t)}{2} f\left(\frac{1}{\rho(x,t)}\right) \nonumber \\
&+& \frac{\rho(x,t)}{2} f\left(\frac{1}{\rho(x+1/\rho,t)}\right) . \qquad
\label{using1}
\end{eqnarray}
In summary, the macroscopic velocity equation corresponding to the optimal velocity model
corresponds to\footnote{If another smoothing function is applied, the last term of Eq. (\ref{finaleq}) is replaced by a similar weighted mean value, as Eq. (\ref{reve}) reveals, but the essence stays the same. That is, the way of looking at the microscopic equations (i.e. the way of defining the density and velocity moments) potentially has some influence on the dynamics, but it is expected to be small.}
\begin{eqnarray}
& & \frac{\partial V(x,t)}{\partial t} + V(x,t) \frac{\partial V(x,t)}{\partial x}  \nonumber \\
&=& - \frac{1}{\rho(x,t)}\frac{\partial }{\partial x} \big[ \rho(x,t)\theta(x,t)\big]
+ \frac{v^0 - V(x,t)}{\tau} \qquad \nonumber \\
&+& \frac{1}{2} f\left(\frac{1}{\rho(x,t)}\right) + \frac{1}{2} f\left(\frac{1}{\rho(x+1/\rho,t)}\right) \, .
\label{finaleq}
\end{eqnarray}
It should be noted that this equation is non-local due to the dependence on $x+1/\rho(x,t)$.
This reflects the anticipatory behavior of drivers, who react to the traffic situation {\it ahead} of them.
From the point of view of traffic simulation, the non-locality does not constitute a problem.
Non-local traffic models such as the gas-kinetic based traffic model summarized in Appendix \ref{gkt} can be even numerically more efficient than local ones with diffusion terms,
that would result from a gradient expansion. In fact, the reason for the numerical inefficiency of explicit solvers for partial differential equations is the diffusion instability, which must be avoided by small time discretizations \cite{numerics}. As pointed out by Daganzo \cite{Daganzo}, a diffusion term also implies theoretical inconsistencies such as the occurence of negative velocities at the end of jam fronts. Therefore, it should be underlined that numerical inefficiencies and theoretical inconsistencies can be avoided by working with the non-local velocity equation rather than with the gradient expansion of it, which will be looked at in the next section.

\subsection{Comparison with Other Macroscopic Traffic Models}

According to the discussion above, a gradient expansion is acceptable in case of small gradients, e.g. when a linear stability analysis is performed. It is also useful to compare different macroscopic traffic models. For this purpose, let us carry out a Taylor approximation of first order. It gives
\begin{eqnarray}
 & & f\left(\frac{1}{\rho(x+1/\rho,t)}\right) \nonumber \\
 &\approx & f\left( \frac{1}{\rho(x,t) + \frac{\partial \rho(x,t)}{\partial x} 
  \frac{1}{\rho(x,t)} }\right) \nonumber \\
 &\approx & f\left( \frac{1}{\rho(x,t)}\left( 1 -  \frac{\partial \rho(x,t)}{\partial x} \frac{1}{\rho(x,t)^2} \right) \right) \nonumber \\
 &\approx& f\left(\frac{1}{\rho(x,t)}\right) + \frac{df(d)}{dd} \cdot \left( - \frac{\partial \rho(x,t)}{\partial x} \frac{1}{\rho(x,t)^3} \right) \, , \quad 
\end{eqnarray}
where we have applied the geometric series expansion $1/(1-z) \approx 1 + z + \dots$
Note that the relation $\rho = 1/d$ and  
\begin{equation}
V_{\rm e}(\rho) =V_{\rm e}\left(\frac{1}{d}\right) 
= v_{\rm o}(d) = v^0 + \tau f(d) = v^0 + \tau f\left(\frac{1}{\rho} \right)
\end{equation}
imply
\begin{eqnarray}
\frac{df(d)}{dd} &=& \left( \frac{d}{d\rho} \frac{V_{\rm e}(\rho) - v^0}{\tau} \right) \frac{d\rho}{dd}
 = \frac{1}{\tau} \frac{dV_{\rm e}(\rho)}{d\rho} \cdot \left(- \frac{1}{d^2}\right) \nonumber \\
 &=& - \frac{\rho^2}{\tau} \frac{dV_{\rm e}(\rho)}{d\rho} \, .
\end{eqnarray} 
Therefore, using Eq. (\ref{using1}), we finally obtain:
\begin{equation}
 \sum_i s(x_i(t)-x) f(t) \approx \rho(x,t) f\left(\frac{1}{\rho(x,t)}\right) 
+ \frac{1}{2\tau} \frac{dV_{\rm e}(\rho)}{d\rho} \frac{\partial \rho(x,t)}{\partial x} \, . 
\end{equation}
Considering $V_{\rm e}(\rho) = v^0 + \tau f(\rho)$ and defining the ``traffic pressure'' as
\begin{equation}
 P(x,t) = \rho(x,t)\theta(x,t) + \frac{v^0 - V_{\rm e}(\rho)}{2\tau} \, ,
 \label{traffpress}
\end{equation} 
the corresponding macroscopic velocity equation becomes
\begin{eqnarray}
& & \frac{\partial V(x,t)}{\partial t} + V(x,t) \frac{\partial V(x,t)}{\partial x}  \nonumber \\
&=& - \frac{1}{\rho(x,t)}\frac{\partial P(x,t)}{\partial x} 
+ \frac{V_{\rm e}(\rho) - V(x,t)}{\tau} \, .
\label{maceqs}
\end{eqnarray}
If the velocity variance $\theta$ is zero, this model corresponds exactly to Payne's macroscopic traffic model with the pressure term \cite{Payne1,Payne2}
\begin{equation}
P(\rho) = \frac{V^0 - V_{\rm e}(\rho)}{2\tau} \, .
\label{Paynepress}
\end{equation}
It should be noted that the gradient $\partial P/\partial x = [dP(\rho)/d\rho]$ $\cdot  \partial \rho/\partial x$ of this pressure becomes zero, whenever the density becomes zero or maximum, as the derivative of\linebreak $dV_{\rm e}(\rho)/d\rho$ vanishes in these situations.

\subsubsection{The Macroscopic Traffic Model by Aw and Rascle}

Note that Daganzo has seriously criticized macroscopic traffic equations
of the type (\ref{maceqs}) \cite{Rascle}. For example, he considered the case of a vehicle queue of
maximum density $\rho = \rho_{\rm jam}$ and speed $V = V_{\rm e}(\rho_{\rm jam}) = 0$, the end of which was assumed to be at some location $x=x_0$. Then,  for the last vehicle in the queue, Eq. (\ref{maceqs}) predicts $V=0$ and $dV/dt = \partial V/\partial t + V\partial V/\partial x < 0$, i.e. the occurence of negative velocities, if pressure relations such as $P = \rho \theta_0 - \eta_0 \partial V/\partial x$ with non-negative parameters $\theta_0$ and $\eta_0$ are assumed \cite{Kerner93}, as it was common at the time when Ref. \cite{Daganzo} was published. 
\par
In order to overcome Daganzo's criticism, Aw and Rascle \cite{Rascle} have proposed the macroscopic velocity equation
\begin{equation}
\frac{\partial}{\partial t} [V + p(\rho)] + V\frac{\partial}{\partial x} [V + p(\rho)] = 0
\end{equation}
with $p(\rho) = \rho^\gamma$. Let us study, how this model relates to the previous macroscopic models.
For this purpose, let us apply the chain rule of Calculus to obtain 
\begin{eqnarray}
 & & \frac{\partial V(x,t)}{\partial t} + V(x,t) \frac{\partial V(x,t)}{\partial x} \nonumber \\
 &=& - \frac{dp(\rho)}{d\rho} \frac{\partial \rho(x,t)}{\partial t} 
  - V(x,t) \frac{dp(\rho)}{d\rho} \frac{\partial \rho(x,t)}{\partial x} \, . \qquad 
\end{eqnarray}
Inserting the continuity equation (\ref{bis}) for $\partial \rho/\partial t$ on the right-hand side,
we get
\begin{eqnarray}
 & & \frac{\partial V(x,t)}{\partial t} + V(x,t) \frac{\partial V(x,t)}{\partial x} \nonumber \\
  &=& \frac{dp(\rho)}{d\rho} \frac{\partial}{\partial x} \big[ \rho(x,t) V(x,t)\big] 
  - V(x,t) \frac{dp(\rho)}{d\rho} \frac{\partial \rho(x,t)}{\partial x} \nonumber \\
   &=& \rho(x,t) \frac{dp(\rho)}{d\rho} \frac{\partial V(x,t)}{\partial x}  \, . 
\end{eqnarray}
By comparison with the macroscopic velocity equation (\ref{maceqs}) we see that the
model by Aw and Rascle does not have a relaxation term $[V_{\rm e}(\rho) - V(x,t)]/\tau$, which would correspond to the limit $\tau \rightarrow \infty$. Moreover, we find
\begin{equation}
 - \frac{1}{\rho} \frac{\partial P(x,t)}{\partial x} = \rho(x,t) \frac{dp(\rho)}{d\rho} \frac{\partial V(x,t)}{\partial x} \, . \label{fit}
\end{equation}
Therefore, the traffic pressure according to the model of Aw and Rascle is a function of the velocity gradient rather than the density gradient, in contrast to Payne's pressure term (\ref{Paynepress}). Consequently, Aw's and Rascle's pressure term must result in a different way than Payne's one. In order to demonstrate this, let us now discuss a generalization of the optimal velocity model and its macroscopic counterpart.

\subsubsection{Non-Local Macroscopic Traffic Models}

It is well-known that the optimal velocity model may produce accidents, if the initial condition,
the optimal velocity function $v_{\rm o}(d)$, and the parameter $\tau$ are not carefully chosen.
In order to have both, the emergence of traffic jams and the avoidance of accidents, we need to assume
that the repulsive interaction force among vehicles does not only depend on the vehicle distance
$d_i(t) = x_{i-1}(t) - x_i(t)$, but also on the vehicle velocity $v_i(t)$ (to reflect the dependence of the safe distance on the vehicle speed) or on the relative velocity 
\begin{equation}
\Delta v_i(t) = v_i(t) - v_{i-1}(t) = - \frac{dd_i}{dt} \, .
\end{equation}
The corresponding generalization of the acceleration equation (\ref{acceq}) reads
\begin{equation}
 \frac{dv_i}{dt} = a_i(t) = \frac{v^0 - v_i(t)}{\tau} + f \Big(d_i(t) , v_i(t) , \Delta v_i(t)\Big) \, .
\end{equation}
This also changes the associated macroscopic traffic equation. Namely, equation (\ref{finaleq})) has to be replaced by
\begin{eqnarray}
& & \frac{\partial V(x,t)}{\partial t} + V(x,t) \frac{\partial V(x,t)}{\partial x} \nonumber \\  
&=& - \frac{1}{\rho(x,t)}\frac{\partial }{\partial x} \big[ \rho(x,t)\theta(x,t)\big]
+ \frac{v^0 - V(x,t)}{\tau} \nonumber \\
&+& \frac{1}{2} f\left(\frac{1}{\rho(x,t)}, V(x,t),\Delta V(x,t) \right) \nonumber \\
&+& \frac{1}{2} f\left(\frac{1}{\rho(x+1/\rho,t)}, V(x+1/\rho,t), \Delta V(x+1/\rho,t)\right) \, . \nonumber \\
& & 
\label{generaleq}
\end{eqnarray}
For the sake of comparison with other macroscopic traffic models and linear stability analyses, let us perform a Taylor approximation of this. First, we may write
\begin{eqnarray}
& & f\left(\frac{1}{\rho(x+1/\rho,t)}, \Delta V(x+1/\rho,t), V(x+1/\rho,t)\right) \nonumber \\
&\approx & f\left(\frac{1}{\rho(x,t)}, \Delta V(x,t), V(x,t)\right) \nonumber \\
&+& \frac{\partial f}{\partial d}  \frac{dd}{d\rho} \Big[ \rho(x+1/\rho,t) - \rho(x,t)\Big] \nonumber \\
&+&  \frac{\partial f}{\partial v}  \Big[ V(x+1/\rho,t) - V(x,t)  \Big] \nonumber \\
&+& \frac{\partial f}{\partial \Delta v}  \Big[ \Delta V(x+1/\rho,t) - \Delta V(x,t)  \Big] \, . 
\end{eqnarray}
Then, we may insert $dd/d\rho = -1/\rho^2$,
\begin{equation} 
 \rho(x+1/\rho,t) - \rho(x,t) \approx \frac{\partial \rho}{\partial x} \,  \frac{1}{\rho}\, , 
\end{equation}
\and
\begin{equation}
 V(x+1/\rho,t) - V(x,t) \approx \frac{\partial V}{\partial x}  \,  \frac{1}{\rho} \, .
\end{equation}
Furthermore, considering $\Delta v_i(t) = - dd_i/dt$, $\rho(x,t) = 1/d_i(t)$, and the continuity equation $d\rho/dt = \partial \rho/\partial t + V\partial \rho/\partial x = - \rho \, \partial V/\partial x$,
we get 
\begin{eqnarray}
\Delta V(x,t) &=& - \frac{d}{dt} \left( \frac{1}{\rho(x,t)}\right) 
= \frac{1}{ \rho(x,t)^2} \, \frac{d\rho(x,t)}{dt} \nonumber \\
&=&  - \frac{1}{\rho(x,t)} \, \frac{\partial V(x,t)}{\partial x} \nonumber \\
&\approx & V(x,t) - V(x+1/\rho,t) 
\end{eqnarray}
and 
\begin{eqnarray}
& & \Delta V(x+1/\rho,t) - \Delta V(x,t) \nonumber \\
&\approx & \frac{\partial \Delta V}{\partial x} \, \frac{1}{\rho} 
\approx - \frac{1}{\rho} \frac{\partial}{\partial x} \left( \frac{\partial V}{\partial x} \frac{1}{\rho} \right) 
\nonumber \\
&=& \frac{1}{\rho^3} \, \frac{\partial \rho}{\partial x} \frac{\partial V}{\partial x} 
- \frac{1}{\rho^2} \frac{\partial^2 V}{\partial x^2} \approx - \frac{1}{\rho^2} \frac{\partial^2 V}{\partial x^2} \, ,
\end{eqnarray}
as a linearization drops products of gradient terms such as $(\partial \rho/\partial x)(\partial V/\partial x)$ (which are assumed to be smaller than the linear terms). Altogether, with $dd/d\rho = -1/\rho^2$ we can write
\begin{eqnarray}
& & f\left(\frac{1}{\rho(x+1/\rho,t)}, V(x+1/\rho,t), \Delta V(x+1/\rho,t)\right) \nonumber \\
&\approx & f\left(\frac{1}{\rho(x,t)}, \Delta V(x,t), V(x,t)\right)
- \frac{1}{\rho^3} \, \frac{\partial f}{\partial d} \frac{\partial \rho}{\partial x} \nonumber \\
& &  \quad +  \frac{1}{\rho} \, \frac{\partial f}{\partial v} \frac{\partial V}{\partial x} 
 - \frac{1}{\rho^2} \, \frac{\partial f}{\partial \Delta v}  \frac{\partial^2 V}{\partial x^2} \, .
\end{eqnarray}
With the definition 
\begin{equation}
V_{\rm o}(\rho, V,\Delta V) = v^0 + \tau f\left(\frac{1}{\rho}, V,\Delta V\right) \, , 
\end{equation}
we may finally write 
\begin{eqnarray}
\frac{\partial V(x,t)}{\partial t} &+& V(x,t) \frac{\partial V(x,t)}{\partial x}  
= - \frac{1}{\rho}\frac{\partial }{\partial x} \big[ \rho(x,t)\theta(x,t)\big] \nonumber \\
&+& \frac{V_{\rm o}(\rho,V,\Delta V) - V(x,t)}{\tau} 
- \frac{1}{2\rho^3} \, \frac{\partial f}{\partial d} \frac{\partial \rho}{\partial x} \nonumber \\
&+&  \frac{1}{2\rho} \, \frac{\partial f}{\partial v} \frac{\partial V}{\partial x} 
-  \frac{1}{2\rho^2} \, \frac{\partial f}{\partial \Delta v}  \frac{\partial^2 V}{\partial x^2} \, .
\end{eqnarray}
Furthermore, let us assume that the variance can be approximated as a function of the density and the average velocity:
\begin{equation}
\theta(x,t) = \theta_{\rm e}\big( \rho(x,t),V(x,t)\big) \, .
\end{equation}
With the definitions
\begin{eqnarray}
\frac{\partial P_1}{\partial \rho} &=& \theta_{\rm e}(\rho,V) + \rho \, \frac{\partial \theta_{\rm e}(\rho,V)}{\partial \rho} + \frac{1}{2\rho^2} \, \frac{\partial f(1/\rho,V,\Delta V)}{\partial d} \, , \qquad \\
\frac{\partial P_2}{\partial V} &=&  \rho \, \frac{\partial \theta_{\rm e}(\rho,V)}{\partial V} 
- \frac{1}{2} \, \frac{\partial f(1/\rho,V,\Delta V)}{\partial v} \, , \\
\eta &=& -  \frac{1}{2\rho^2} \, \frac{\partial f(1/\rho,V,\Delta V)}{\partial \Delta v} 
\end{eqnarray}
(where $\eta$ should be greater than zero),
we may also write the linearized macroscopic traffic equations as
\begin{eqnarray}
& & 
\frac{\partial V(x,t)}{\partial t} + V(x,t) \frac{\partial V(x,t)}{\partial x}  \nonumber \\
&=& - \frac{1}{\rho}\frac{\partial P_1}{\partial \rho} \frac{\partial \rho}{\partial x} 
- \frac{1}{\rho}\frac{\partial P_2}{\partial V} \frac{\partial V}{\partial x} 
+ \eta  \frac{\partial^2 V}{\partial x^2}\nonumber \\ 
&+& \frac{V_{\rm o}(\rho,\Delta V,V) - V(x,t)}{\tau} \, .
\label{put}
\end{eqnarray}
The term $\eta \partial^2 V/\partial x^2$ can be interpreted as viscosity term and has some smoothing effect. Further viscosity (and diffusion) terms may be derived by second-order Taylor expansions. 
\par
Note that the pressure term $P_2$ looks similar to Eq. (\ref{fit}). Therefore, it should be possible to derive a pressure term corresponding Aw's and Rascle's model from a suitable microscopic traffic model, but one would expect additional terms such as (\ref{traffpress}) to occur as well.

\section{Summary, Discussion, and Conclusions}

In this paper, we have discussed several approaches to derive macroscopic traffic equations from microscopic car-following models. It has been pointed out that a Taylor approximation should be used only for linear stability analyses, as the gradients may otherwise be too large for the approximation to work. Further undesireable consequence of a gradient expansion are the possible occurence of negative velocities, diffusion instabilities, and inefficient numerical solution methods. 
\par
The linear interpolation approach often works well in practise, but it is theoretically inconsistent as it violates the continuity equation which is required for the conservation of the vehicle number. In constrast, the smooth particle hydrodynamics approach was suited in all respects. It led to a non-local macroscopic traffic model, as did the gas-kinetic based traffic model. In order to have a realistic traffic dynamics (in particular accident avoidance if a vehicle with speed $v^0$ approaches a standing car), one needs to take into account that the repulsive vehicle interactions depend not only on the vehicle distance, but also on the relative velocity and the vehicle velocity. This leads to a specification of the traffic pressure which contains variance-dependent terms, additional
terms proportional to $\partial \rho/\partial x$ as in Payne's model, and further terms proportional to $\partial V/\partial x$ as in Aw's and Rascle's model. While the variance-dependent term describes dispersion effects, Payne's, Aw's and Rascle's terms reflect effects of vehicle interactions. 
Finally note that, in case of multi-lane traffic, the additional inter-lane variance
\begin{equation}
 \Theta(x,t) = \frac{1}{L} \sum_{l=1}^L \frac{\rho_l(x,t)}{\rho(x,t)} [V_l(x,t) - V(x,t)]^2 \, , 
\end{equation}
must be added to the inner-lane variance $\theta(x,t)$,
where $\rho_l(x,t)$ is the density and $V_l(x,t)$ the average velocity in lane $l$ at location $x$ and time $t$ \cite{Review,Shvetsov}.

\begin{acknowledgement}
The author would like to thank for the inspiring discussions with the participants of the Workshop on ``Multiscale Problems and Models in Traffic Flow'' organized by Michel Rascle and Christian Schmeiser at the Wolfgang Pauli Institute in Vienna from May 5--9, 2008, with partial support by the CNRS.
\end{acknowledgement}
\appendix

\section{The Non-Local, Gas-Kinetic Based Traffic Model} \label{gkt}

For comparison, let us shortly recall the form of the non-local gas-kinetic based traffic model (GKT model). This has been derived via a collision approximation, see Ref. \cite{GKT}. This can be written in the form of equation (\ref{maceqs}) with $P(x,t) = \rho(x,t)\theta(x,t)$,
but $V_{\rm e}(\rho)$ must be replaced by a non-local expression 
\begin{equation}
 V_{\rm g}(\rho,V,\theta,\rho_{_+},V_{_+},\theta_{_+} )
 = v^0 \underbrace{-\tau [1-p(\rho_{_+})]\chi(\rho_{_+}) \rho_{_+} 
 B(\Delta)}_{\rm repulsive\ interaction\ term} \, . 
\label{eqVderiv}
\end{equation}
Here, the index ``+'' indicates evaluation at the advanced ``interaction point'' $x+s_0+TV$,
where $s_0$ represents the minimum vehicle distance and $TV$ the velocity-dependent safety distance. The related non-locality has some effects, which other macroscopic
models generate by their pressure and viscosity terms. 
The dependence of the non-local repulsive interaction on
the effective dimensionless velocity difference 
\begin{equation}
 \Delta = \frac{V-V_{_+}}{\sqrt{\theta - 2 r \sqrt{\theta \theta_{_+}} + \theta_{_+}}} 
\end{equation}
takes into account effects of the velocity variances $\theta$, $\theta_{_+}$,
and velocity correlations $r$ among successive cars \cite{Shvetsov}.
Furthermore, the ``Boltzmann factor''
\begin{equation}  
B(\Delta) =  \Big( \theta - 2 r \sqrt{\theta \theta_{_+}} + \theta_{_+} \Big) \Big[ 
    \Delta N(\Delta) 
           + \big(1+\Delta^2\big) E(\Delta) \Big] 
\label{boltzfact}
\end{equation}
in the braking term is monotonically increasing with $\Delta V$. It
contains the normal distribution
\begin{equation}
 N(\Delta) = \frac{\mbox{e}^{-\Delta^2/2}}{\sqrt{2\pi}}
\end{equation}
and the Gaussian error function 
\begin{equation}
 E(\Delta) = \int\limits_{-\infty}^{\Delta} dz \, N(z) . 
\end{equation}
To close the system of equations, the velocity 
correlation $r$ is specified as a function of the density in accordance with empirical observations. 
Moreover,  
for a description of the presently known properties
of traffic flows it seems sufficient to set
\begin{equation}
 \theta = A(\rho) V^2 \, . 
\label{th1}
\end{equation}
This guarantees that the velocity variance will vanish whenever the average
velocity goes to zero, but it will be positive otherwise. It should be noted that
the variance prefactor $A$ is higher in congested traffic than in free traffic. 
The {\em ``effective cross section''} is, finally, specified via
\begin{equation}
 [1-p(\rho)]\chi(\rho) = \frac{v^0 \rho T^2}{\tau A(\rho_{\rm jam}) 
 (1-\rho/\rho_{\rm jam})^2} \, ,
\label{bleibt}
\end{equation}
where $T$ is the safe time headway and $\rho_{\rm jam}$ the maximum vehicle density.
This formula makes also sense in the low-density limit $\rho \to 0$, where
$\chi \to 1$ and $p\to 1$. 
\par
A linear stability analysis of the
non-local traffic model can be done via a gradient expansion. It results in equations of the kind (\ref{put}) and further viscosity and diffusion terms \cite{GKTexpans}.

\end{document}